\documentclass[reprint,aps,prc,amsmath,amssymb,nofootinbib,superscriptaddress]{revtex4-2}

\usepackage{braket} 
\usepackage[dvipsnames]{xcolor}  
\usepackage{float}
\usepackage[T1]{fontenc}
\usepackage[utf8]{inputenc}
\usepackage{graphicx,color,rotating,pifont}
\usepackage{amsmath,amssymb}
\usepackage{mathrsfs}
\usepackage{mathtools}
\usepackage{subfigure}
\usepackage{siunitx}
 
\usepackage{bm}
\usepackage{ae}
\usepackage{dcolumn}
\usepackage{txfonts}
\usepackage{tensor}
\usepackage{braket}
\usepackage{booktabs}
\usepackage{xspace}
\usepackage{soul}
\usepackage{multirow}
\usepackage{tikz}
\usepackage{pgffor} 

\usepackage[normalem]{ulem}

\setstcolor{red}
\setul{}{.2ex} 

\usepackage{shortcuts}

\usepackage{mathtools} 

\usepackage{hyperref}
\usepackage{cleveref}

\usepackage{xspace}
\makeatletter
\DeclareRobustCommand\onedot{\futurelet\@let@token\@onedot}
\def\@onedot{\ifx\@let@token.\else.\null\fi\xspace}

\makeatother

\begin{document}
  
\title{Benchmarking projected generator coordinate method for nuclear Gamow–Teller transitions}

 
 \author{R. N. Chen} 
  \affiliation{School of Physics and Astronomy, Sun Yat-sen University, Zhuhai 519082, P.R. China}   
 \affiliation{Guangdong Provincial Key Laboratory of Quantum Metrology and Sensing, Sun Yat-Sen University, Zhuhai 519082, P.R. China }

 \author{X. Lian} 
  \affiliation{College of Physics, Sichuan University, Chengdu 610065, China}  
 
 \author{J. M. Yao}  
 \email{Contact author: yaojm8@sysu.edu.cn}
  \affiliation{School of Physics and Astronomy, Sun Yat-sen University, Zhuhai 519082, P.R. China}   
 \affiliation{Guangdong Provincial Key Laboratory of Quantum Metrology and Sensing, Sun Yat-Sen University, Zhuhai 519082, P.R. China } 
 
 \author{C. L. Bai} 
 \email{Contact author: bclphy@scu.edu.cn}
  \affiliation{College of Physics, Sichuan University, Chengdu 610065, China}

\date{\today}

\begin{abstract}
\begin{description}

\item[Background] 
The quantum-number projected generator coordinate method (PGCM) has gained increasing attention, in part due to its combination with the in-medium similarity renormalization group (IMSRG) to describe collective excitations in medium-mass deformed nuclei, as well as  nuclear matrix elements (NMEs) of neutrinoless double-beta ($0\nu\beta\beta$) decay. 

\item[Purpose] Extending the PGCM to  Gamow–Teller (GT) transitions and two-neutrino double-$\beta$ ($2\nu\beta\beta$) decay is nontrivial, as it requires an accurate description of not only nuclear ground states but also a large number of excited states.  In this work, we aim to achieve a minimal extension of the PGCM to describe GT transition strengths in even-even nuclei and to compute the NME of $2\nu\beta\beta$ decay  .   
 
\item[Method] Within the PGCM framework, the wave functions of odd–odd nuclei are constructed as superpositions of neutron and proton quasiparticle configurations built on quasiparticle vacua constrained to have, on average, odd neutron and odd proton particle numbers. The angular momentum and particle numbers associated with the underlying mean-field states are restored through projection techniques. Using a shell-model Hamiltonian defined in the 
$fp$ shell, we assess the validity of this approach by benchmarking GT transitions in calcium and titanium isotopes, as well as the  $2\nu\beta\beta$ decay of 
$^{48}$Ca to $^{48}$Ti, against exact solutions. For comparison, we also confront our results with those obtained from configuration-interaction calculations employing different particle–hole truncation schemes, both with and without IMSRG evolution.

\item[Results] The PGCM generally reproduces GT transitions to both  low-lying  and giant   resonance states in $^{42\text{–}48}$Ca and $^{42\text{–}48}$Ti. Although the agreement with exact results deteriorates as the number of valence nucleons increases, the overall description remains robust. The NME of the $2\nu\beta\beta$ decay from $^{48}$Ca to $^{48}$Ti is calculated without invoking the closure approximation. We find that the PGCM overestimates this matrix element by about 57\%, primarily due to an overestimation of the GT transition strength from $^{48}$Ti to the first excited state of $^{48}$Sc. Overall, the performance of the PGCM is comparable to, and in some cases exceeds, that of the CI calculation with the two-particle–two-hole truncation for the nuclei concerned.

\item[Conclusions] The currently implemented PGCM framework provides a reliable description of GT transitions from even–even nuclei to low-lying states of odd–odd nuclei in regions not far from closed shells. As the number of valence nucleons increases, deviations from exact results become more noticeable, reflecting the growing importance of complex many-body correlations. These discrepancies are expected to be reduced by extending the set of generator coordinates and by incorporating the IMSRG evolution, which together offer a  promising route toward further enhancing the predictive power of the PGCM framework.  
 
\end{description}
\end{abstract}

\pacs{Valid PACS appear here}
\maketitle
\section{Introduction}
Nuclear weak processes, including single-$\beta$ and double-$\beta$ decays, are a central topic of interdisciplinary research in nuclear physics, particle physics, and astrophysics. They play a central role in elucidating nuclear stability, the nucleosynthesis of elements in the universe~\cite{Langanke:2003RMP,Fischer:2024PPNP,Suzuki:2022PPNP,Hu:2025}, and in probing physics beyond the Standard Model~\cite{Herczeg:2001PPNP,Severijns:2006,Otten:2008RPP,Falkowski:2021JHEP}. In this context, accurate nuclear matrix elements (NMEs) are indispensable, as they provide the critical link between measured observables and the underlying physics~\cite{RevModPhys.80.481,Engel_2017,universe6120225}. For example, the NME governing superallowed $0^+\to 0^+$ Fermi transitions connects the measured half-life to the $V_{ud}$ element of the Cabibbo–Kobayashi–Maskawa (CKM) matrix~\cite{Hardy:2020,Severijns:2023}, while the NME for neutrinoless double-$\beta$ ($0\nu\beta\beta$) decay relates the decay half-life to the effective neutrino mass within the standard mechanism~\cite{Agostini:2023}. Because NMEs cannot be accessed directly by experiment, they must be determined through nuclear-structure calculations, which require accurate wave functions for both initial and final states, together with reliable transition operators. Achieving such precision remains a formidable challenge, owing to the complexity of the nuclear force and the quantum many-body problem, whose exact solution rapidly becomes intractable as the number of nucleons increases.

Nuclear weak processes are typically modeled using a variety of nuclear-structure approaches that rely on different levels of approximation. A prominent example is the $0\nu\beta\beta$ decay, for which different methods predict a wide spread in the corresponding NMEs, resulting in uncertainties of a factor of $2$–$3$ or even larger~\cite{Engel_2017,Yao:2022PPNP}. Because each model is built upon distinct approximations and effective nuclear interactions, systematically reducing these discrepancies remains highly challenging.
In recent years, significant progress has been achieved by combining {\em ab initio} nuclear methods ~\cite{Hergert:2020} with conventional nuclear many-body solvers, enabling the calculation of NMEs of weak processes using transition operators derived from chiral effective field theory~\cite{Weinberg:1991}. Notable examples include the valence-space in-medium similarity renormalization group (IMSRG)~\cite{Stroberg:2019ARNPS} and the in-medium generator coordinate method (IM-GCM)~\cite{Yao:2018PRC}, both of which have been successfully applied to $0\nu\beta\beta$ decay in candidate nuclei~\cite{Yao:2020PRL,Belley:2020ejd,Belley:2024PRL,Belley:2023_all}.

It is worth noting that the IM-GCM is a combination of multi-reference IMSRG~\cite{Hergert:2014,Yao:2018PRC} and quantum-number projected GCM (PGCM)~\cite{Ring:1980}. The former is used to capture dynamic correlations associated with high-energy particle-hole excitations, while the latter is employed to include the collective (static) correlations associated with pairing and deformation. However, extending the IM-GCM to compute NMEs for single-$\beta$ decay and two-neutrino double-$\beta$ ($2\nu\beta\beta$) decay is considerably more challenging than for $0\nu\beta\beta$ decay. These processes require an accurate description not only of the nuclear ground states but also of a large number of excited states, all of which may contribute significantly to the decay. This difficulty is particularly pronounced for GT transitions of even–even nuclei, where many intermediate states of the corresponding odd–odd nuclei must be treated explicitly. In such systems, numerous configurations are nearly degenerate and must be incorporated into the configuration-mixing calculation, posing substantial challenges for the PGCM framework.

In this work, we generalize the PGCM framework to odd–odd nuclear systems and apply this extended framework to investigate GT transitions in even–even Ca and Ti isotopes and to evaluate the NME of the $2\nu\beta\beta$ decay of $^{48}$Ca. The wave functions of the even–even nuclei are constructed as superpositions of quantum-number–projected Hartree–Fock–Bogoliubov (HFB) states with different intrinsic quadrupole deformations, whereas those of the odd–odd nuclei are approximated as superpositions of quantum-number–projected two-quasiparticle configurations built on a HFB reference state. 
We benchmark this framework  using a shell-model Hamiltonian defined within the $fp$ shell for which exact solutions are available. To elucidate the origin of the discrepancies between the PGCM results and the exact solutions, we further compare our calculations with configuration-interaction (CI) results that include particle–hole excitations of different truncation levels, with and without the IMSRG evolution for the interaction and GT transition operators. 

It is worth noting that several related frameworks have been applied to the study of GT transitions. For example, the performance of angular-momentum projected Hartree-Fock (PHF) with projection after variation  for excitation-energy spectra has been benchmarked against full CI calculations, demonstrating that quantitative agreement can already be achieved at the PHF level~\cite{Lauber:2021}. In addition, both variation-after-projection and GCM within the antisymmetrized molecular dynamics framework  have been applied to describe GT transitions from the ground state of $^{14}$N to low-lying states of $^{14}$C~\cite{Kanada-Enyo:2014}. Isospin projection has also been implemented in combination with the PGCM for nuclear $\beta$ decays~\cite{Konieczka:2016,Morita:2018}. More recently, a density functional theory-based no-core CI framework has been extended to describe GT transitions and the NME of the $2\nu\beta\beta$ decay in $^{48}$Ca~\cite{Miskiewicz:2025}.

This paper is organized as follows. In Sec.~\ref{sec:framework}, we introduce the PGCM and CI frameworks based on a shell-model Hamiltonian. Section~\ref{sec:Results and discussion} presents and discusses the GT transition strengths of even–even Ca and Ti isotopes obtained with both methods, in comparison with exact solutions, and also reports the NME of the $2\nu\beta\beta$ decay in $^{48}$Ca. Finally, Sec.~\ref{sec:summary} summarizes the main conclusions and outlines future perspectives.

 \section{The methods}
 \label{sec:framework}
\subsection{The Hamiltonian}
The employed shell-model Hamiltonian is written in the second quantization form,
\beqn
H&=&H_m -  \frac{1}{4}\sum_{pqrsJM} {\cal V}^J_{pqrs} 
 (-1)^{J+M}[c^\dagger_pc^\dagger_q]_{JM}[\tilde c_r \tilde c_s]_{J-M}, 
\eeqn 
where the first term $H_m$, called an “unperturbed”  term, is given by the sum of single-particle energies $\varepsilon_p$,   
 \beq
 H_{m} = \sum_{p} \varepsilon_p c^\dagger_p c_p,
 \eeq
 and the second term for the two-body residual interaction. The indices $p$ (also $q, r$, and $s$) specifies the single-particle state of quantum numbers $t_pn_pl_pj_p$, where the 3rd-component of the isospin $t_p$ distinguishes neutron and proton states. The J-coupled two-particle creation and annihilation operators are defined as
\beq
[c^\dagger_pc^\dagger_q]_{JM}
=\sum_{m_{p}, m_{q}}\langle j_{p} m_{p} j_{q} m_{q} \mid J M\rangle c_{j_{p} m_{p}}^{\dagger} c_{j_{q} m_{q}}^{\dagger},
\eeq
and
\beq
[\tilde{c}_r\tilde{c}_s]_{J-M}
=\sum_{m_{r}, m_{s}}\langle j_{r} m_{r} j_{s} m_{s} \mid J -M\rangle \tilde{c}_{j_{r} m_{r}}  \tilde{c}_{j_{s} m_{s}} 
\eeq 
where the two particles $p$ and $q$ ($r$ and $s$) couple to total angular
momentum $J$, and $\tilde{c}_{j m}=(-1)^{j-m}c_{j-m}$. The ${\cal V}^J_{pqrs}$ is the unnormalized two-body interaction matrix element
 \beq
 {\cal V}^J_{pqrs}
 ={\cal N}^{-1}_{pq}(J){\cal N}^{-1}_{rs}(J) \langle pq(J)\vert V \vert rs(J)\rangle
 \eeq
 where the normalization factor reads ${\cal N}_{pq}(J)=\sqrt{1+\delta_{pq}(-1)^J}/(1+\delta_{pq})$, and $\vert rs(J)\rangle$ are antisymmetrized and
normalized two-particle states. The $\varepsilon_p$ and ${\cal V}^J_{pqrs}$
 are free parameters that have been fitted to nuclear low-lying states of a particular mass region~\cite{Caurier:2004gf}.  In this work, we   employ the GXPF1A~\cite{Honma:2004} shell model Hamiltonian, which is defined  within the $fp$ shell, consisting of the $0f_{7/2}, 0f_{5/2}, 1p_{3/2}$, and $1p_{1/2}$ spherical harmonic orbitals for both protons and neutrons on top of an $^{40}$Ca inert core.
 

   \subsection{The projected generator coordinate method}
 In the PGCM, the wave function  of  nuclear state is constructed as a superposition of symmetry-projected HFB wave functions 
\begin{equation}
\label{eq:GCM_wavefunction}
|\Psi^{J_i M_iN_i Z_i}_m\rangle = \sum_{c}
f^{J_i, m}_c \ket{J_i M_i N_iZ_i c}
\end{equation}
where $m$ is a label distinguishing the states with the same spin-parity $J^\pi$. The symbol $c$ is a collective label for $(K, \bm{q})$. The basis states $\ket{J_i M_i N_iZ_ic}$ are constructed as follows 
\begin{equation}
\ket{J_i M_iN_iZ_i c}=\hat{P}^{J_i}_{M_i K} \hat{P}^{N_i} \hat{P}^{Z_i} |\Phi(\bm{q})\rangle.
\end{equation}
Here $\hat{P}^{J_i}_{M_i K_i}$ is the operator that projects the intrinsic wave function $|\Phi(\bm{q})\rangle$ onto components with well defined angular momentum $J_i$, and its projection along the $z$-axis $M_i$, and body-fixed $3$rd axis $K_i$. The operators $\hat{P}^{N_i}$ and $\hat{P}^{Z_i}$
project the wave function onto components with well-defined neutron number $N_i$ and proton
number $Z_i$.  The projection operators produce basis states that are not orthonormal. 

The mixing weight in (\ref{eq:GCM_wavefunction}) is determined from the variational principles which leads
to the Hill-Wheeler-Griffin (HWG) equation~\cite{Ring:1980}
\begin{equation}
\label{eq:HWG}
\sum_{c'} \big[ \mathcal{H}^{J_i}_{cc'}  - E^{J_i}_m
\mathcal{N}^{J_i}_{cc'}  \big] f^{J_i m}_{c'} = 0,
\end{equation}
where the Hamiltonian and norm kernels $\mathcal{H}$ and $\mathcal{N}$ are given by the expressions
\begin{eqnarray}
\mathcal{H}^{J_i}_{cc'}  &=& \bra{\Phi(\bm{q})} \hat{H}\hat{P}^{J_i}_{KK'} \hat{P}^{N_i} \hat{P}^{Z_i} \ket{\Phi(\bm{q}')},\\ 
\mathcal{N}^{J_i}_{cc'}  &=&\bra{\Phi(\bm{q})}  \hat{P}^{J_i}_{KK'} \hat{P}^{N_i} \hat{P}^{Z_i} \ket{\Phi(\bm{q}')},
\label{eq:ovker}
\end{eqnarray}
and $E^{J_i}_{m}$ is the energy of the $m$-th state with angular momentum $J_i$. 
In this work, we only consider the GT transition from the ground state of an even-even nucleus to different states of the odd-odd nucleus. In this case, the spin parity of the ground state for the initial state is $J^{\pi}=0^+$, which simplifies the calculation. 

The daughter nucleus is an odd-odd nucleus, whose wave function  is constructed similarly, with the basis function 
 $\ket{J_f M_f N_fZ_fc}$ constructed as follows 
\begin{equation}
\label{eq:wf_odd_odd}
\ket{J_f M_fN_fZ_f c}=\hat{P}^{J_f}_{M_f K} \hat{P}^{N_f} \hat{P}^{Z_f} A^\dagger_{pn}|\Phi(\bm{q})\rangle,
\end{equation}
  where the symbol $c$ is a collective label for $c\equiv\{K, \mathbf{q}, p, n\}$. $|\Phi(\bm{q})\rangle$ is a quasiparticle vacuum state from the HFB calculation with the average particle numbers constrained to be odd neutron and odd proton.  The $A^\dagger_{pn}$ is a combined operator creating one quasiparticle neutron and one quasiparticle proton, 
  \beq
  A^\dagger_{pn} = \beta^+_p \beta^+_n.
  \eeq
  For the sake of simplicity,  the HFB wave function $|\Phi(\bm{q})\rangle$ is restricted to have axial symmetry. In this case, quasiparticle operators do not have a well-defined total angular momentum.  Instead, each is labeled by $\Omega^\pi$, where $\Omega_{n(p)}$ denotes the projection of the single-quasiparticle angular momentum onto the intrinsic symmetry axis, and $\pi$ represents the parity of the state.  The quasiparticle pair creation operator $A^\dagger_{pn}$ obeys the $K$ selection rule $K_{pn} = \Omega_n + \Omega_p$ and parity selection rule $\pi_n\pi_p  = +1$. We focus on the final states of odd-odd nuclei with spin-parity $1^+$, which involve components with $K_{pn} = 0, \pm 1$. These components are mixed to construct the total wave function of the final state. 
  
  The norm and Hamiltonian kernels in the HWG equation (\ref{eq:HWG}) for the odd-odd nuclei are slightly more complicated than those for the even-even nuclei and they are given by
  \beqn
  N^{J_f}_{cc'} &=& \langle \bm{q} \vert A_{np} \hat P^{J_f}_{KK'} \hat P^{N_f} \hat P^{Z_f}  A^\dagger_{n'p'}\vert \bm{q}'\rangle,\\
  H^{J_f}_{cc'} &=& \langle \bm{q} \vert  A_{np} \hat H\hat P^{J_f}_{KK'} \hat P^{N_f} \hat P^{Z_f}  A^\dagger_{n'p'}\vert \bm{q}'\rangle
  \eeqn
  If the dimension of the quasiparticle states is $M_{qp}$ and the number of deformed configurations is denoted as $N_q$, there are about $M^4_{qp} N^2_q$ kernels to be computed. To simplify the computation,  we truncate the quasiparticle configurations based on the following criterion
    \beq
   E_p + E_n  \leq   E_{\rm cut}.
  \eeq 
  We choose the cut-off value $E_{\rm cut}=48$ MeV, which turns out to be sufficient for giving a convergent solution to the excited states of the odd-odd nuclei in the energy region of interest.  Besides, if not mentioned specifically, only the lowest-energy HFB state is employed for the odd-odd nuclei, i.e., $N_q=1$.  

 We solve the HWG equation (\ref{eq:HWG}) for both even-even and odd-odd nuclei in the standard way \cite{Ring:1980}, by diagonalizing the norm kernel to obtain a basis of ``natural states'' and then diagonalizing the Hamiltonian $H$ in that basis.  The second diagonalization can be numerically unstable, a problem we deal with by truncating the natural basis to include only states with norm eigenvalues larger than a reasonable value.

   \subsection{The configuration-interaction (CI) method with particle-hole truncation} 
   
In the CI method, the nuclear many-body wave function $|\Psi\rangle$ is expanded in a basis of particle--hole excitations built on a reference state as
\begin{equation}
\label{eq:CI_wavefunctions}
|\Psi\rangle
= D_0 |\Phi\rangle
+ \sum_{m,i} D^{m}_{i} |\Phi^{m}_{i}\rangle
+ \sum_{mn,ij} D^{mn}_{ij} |\Phi^{mn}_{ij}\rangle
+ \cdots ,
\end{equation}
where $|\Phi\rangle$ denotes the reference state, which is a  HF state. The states $|\Phi^{m}_{i}\rangle = a^{\dagger}_{m} a_{i} |\Phi\rangle$
represent one-particle--one-hole (1p1h) excitations, with the indices $m$ and $i$ labeling particle and hole single-particle orbitals, respectively. Similarly,
$|\Phi^{mn}_{ij}\rangle = a^{\dagger}_{m} a^{\dagger}_{n} a_{j} a_{i} |\Phi\rangle$
correspond to two-particle--two-hole (2p2h) excitation configurations. The ellipsis indicates higher-order particle--hole excitations.

If all $A$-particle--$A$-hole configurations are included, the CI expansion in Eq.~(\ref{eq:CI_wavefunctions}) yields the exact solution of the many-body problem. In practice, however, the expansion must be truncated at a finite excitation level. The CI($m$p$m$h) scheme, which retains configurations up to $m$-particle--$m$-hole excitations, therefore provides a controlled framework for assessing the role of many-body correlations and for elucidating the origin of discrepancies between PGCM results and exact shell-model calculations.

The many-body Hamiltonian matrix of CI exhibits a block-diagonal structure due to the symmetries of the Hamiltonian. In the M-scheme basis, two symmetries can be exploited: rotational symmetry about the $z$-axis and parity symmetry. The associated symmetry operators are the $z$-projection of angular momentum $J_z$ and the space inversion operator $\Pi$, with corresponding eigenvalues $M$ and $\pi$, respectively. Consequently, only the block matrices specified by fixed values of $M$ and $\pi$ need to be considered in calculations. 
  
  For the ground state of even-even nuclei with quantum numbers $J^{\pi}=0^{+}$, the $M^{\pi}$ subspace must be restricted to $0^{+}$. For excited states of odd-odd nuclei with $J^{\pi}=1^{+}$, both $M^{\pi}=0^{+}$ and $\pm1^{+}$ subspaces must be included. We note that the particle-hole truncation scheme used in this paper does not break rotational invariance in the many-body space. As a result, the calculated energy levels for different $M$ values remain degenerate, making it sufficient to compute the excited states only for $M^{\pi}=1^{+}$. Moreover, we note that only the $f_{7/2}$ orbital is involved in the reference state. Therefore, all excitation configurations within the $f_{7/2}$ orbital are equally important to the reference determinant. Consequently, we do not include particle–hole excitations for configurations among the multiplets of the $f_{7/2}$ orbital.

   \section{Results and discussion}
 \label{sec:Results and discussion}
  \subsection{The GT transition matrix elements}
  The strength for the GT transition from the ground state ($0^+_1$) of an even-even nucleus to the $m$-th excited state ($1^+_m$) is given by~\footnote{Note that the axial-vector coupling constant $g_A$ is not multiplied to the GT transition operator.}
         \beq
      B({\rm GT}^\mp: 0^+_1 \to 1^+_m) = \Big\vert  \langle 1^+_m  \vert \vert \sigma \tau^\mp\vert \vert 0^+_1\rangle \Big\vert^2, 
      \eeq
where $\tau^-$ converts a neutron to a proton.  For the convenience of comparison, we introduce the distribution of strength function as follows, 
      \beq
      S(\rm GT^{\mp},E, \Gamma) = \sum_m  B({\rm GT}^\mp: 0^+_1 \to 1^+_m) {\cal L}(E,  E_{1^+_m}, \Gamma)
      \eeq
 where  the Lorentzian function $L(E, E_{1^+_m}, \Gamma)$ is introduced to smooth the strength function
      \beq
      {\cal L}(E, E_{1^+_m}, \Gamma) =  \dfrac{1}{\pi} \dfrac{\Gamma/2} {(\Gamma/2)^2 + ( E - E_{1^+_m})^2} 
      \eeq
and the width $\Gamma$ is set as 0.5 MeV.

\begin{figure}[]
\centering 
\includegraphics[width=8.5cm]{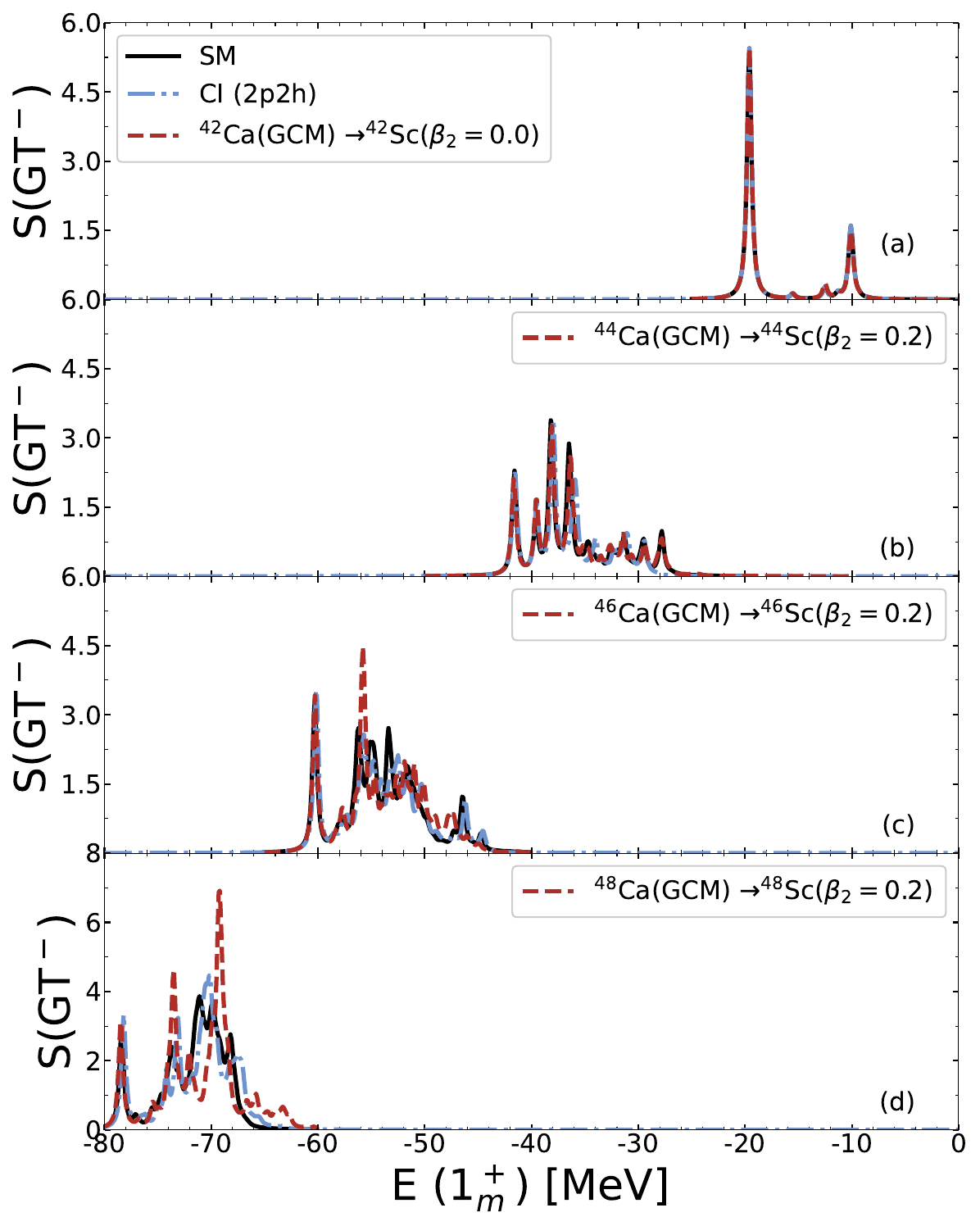}   
\caption{ (Color online)   The distribution of ${\rm GT}^-$ transition strength $B({\rm GT}^-: 0^+_1 \to 1^+_m)$ from $^{42-48}$Ca to $^{42-48}$Sc as a  function of the energy of the $1^+_m$ states in $^{42-48}$Sc isotopes from the PGCM calculation, in comparison with the results of exact shell-model  and  CI(2p2h) calculations. }
\label{fig:Ca-isotopes} 
\end{figure} 

\begin{figure}[]
\centering 
\includegraphics[width=8.5cm]{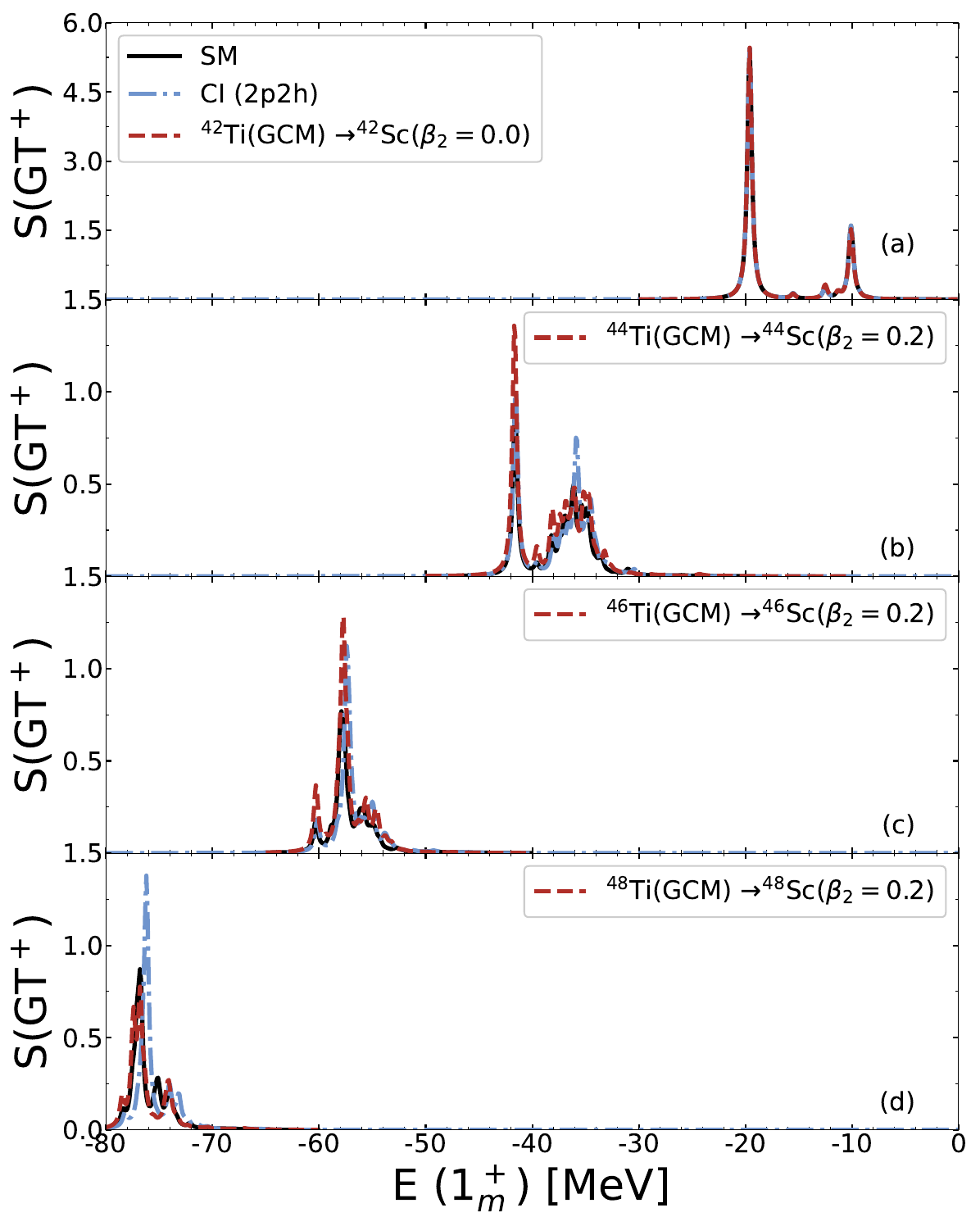}   
\caption{ (Color online) The same as Fig.~\ref{fig:Ca-isotopes}, but for the ${\rm GT}^+$ transitions from $^{42-48}$Ti to $^{42-48}$Sc, respectively.  }
\label{fig:Ti-isotopes} 
\end{figure} 

Figures~\ref{fig:Ca-isotopes} and \ref{fig:Ti-isotopes}  display the strength distributions of the GT$^-$ transition from $^{42-48}$Ca to $^{42-48}$Sc, and from $^{42-48}$Ti to $^{42-48}$Sc, respectively. 
It is shown that generally the PGCM is able to reproduce the results of shell-model calculations quite well for all the cases, even though the discrepancy shows up gradually with the increase of valence nucleons. A similar performance of the CI(2p2h) as that of the PGCM is observed. Quantitatively, the PGCM performs slightly better for the low-lying transitions, but slightly worse for the high-lying states. This difference is probably attributed to the fact that the low-lying states of odd-odd Sc isotopes are dominated by the collective correlations, which can be better captured in the PGCM through the mixing of different deformed configurations. In contrast, some of the important 2p2h excitation configurations are missing in the PGCM and they have a nontrivial contribution to the GT transitions to high-lying states.

   \begin{figure}[]
\centering 
\includegraphics[width=8.5cm]{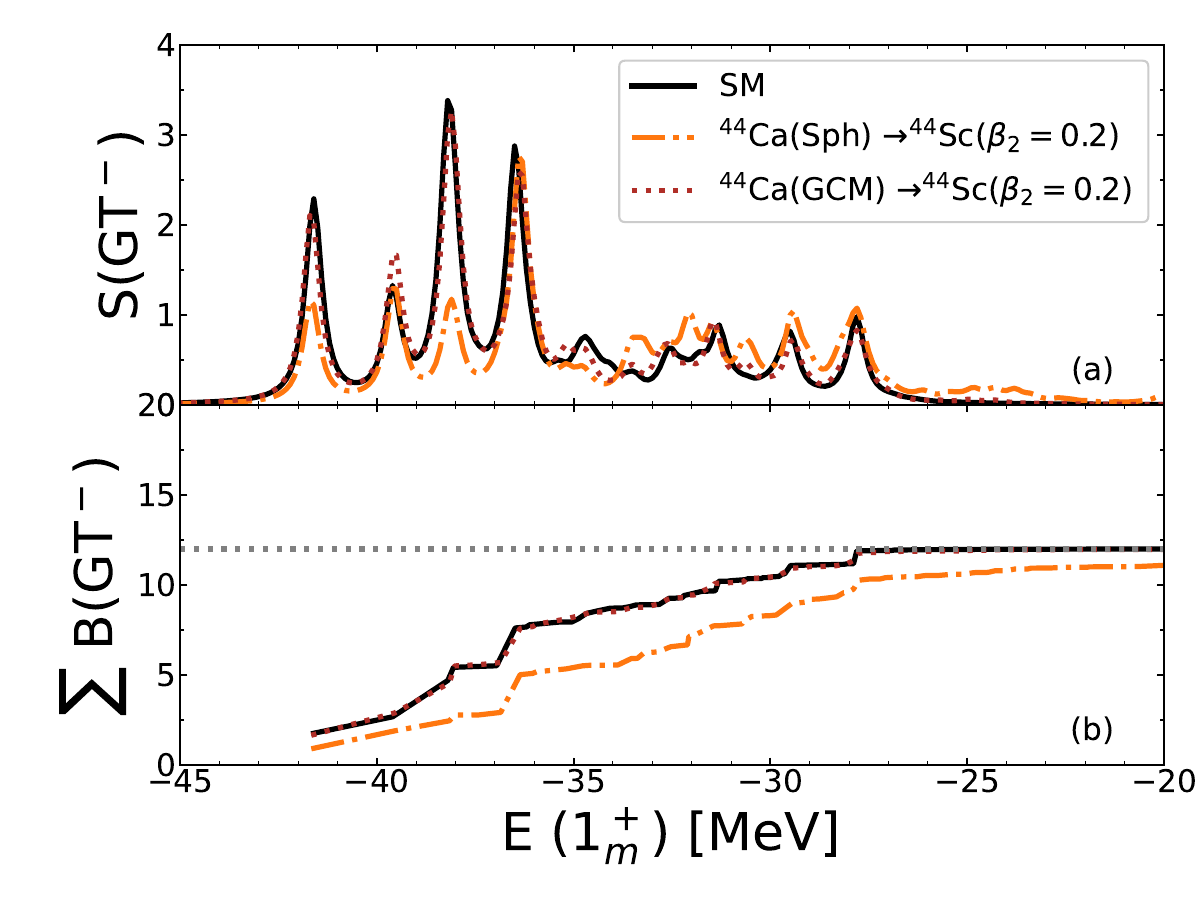}   
\caption{(Color online)  (a) The distribution of ${\rm GT}^-$ transition strength  for $^{44}$Ca as a  function of the energy of the $1^+$ states in $^{44}$Sc. (b) The cumulated GT transition strength. The results of the PGCM calculations with the wave function of $^{44}$Ca constructed using either the pure spherical state or GCM state are compared to the shell-model results.  }
\label{fig:Ca44-sum-rule-compare} 
\end{figure}

To investigate the impact of shape mixing on the GT transition strengths, we compare in Fig.~\ref{fig:Ca44-sum-rule-compare} the results obtained using either a purely spherical state or the GCM ground-state wave function for $^{44}$Ca. The $1^+_m$ states of $^{44}$Sc are approximated by Eq.~(\ref{eq:wf_odd_odd}) with $\mathbf{q}\equiv\{\beta_2=0.2\}$. The use of the GCM wave function for $^{44}$Ca considerably improves the description of the GT transition strengths to both the low-lying and high-lying $1^+$ states in $^{44}$Sc. Moreover, the cumulated GT strength from the exact solution is reproduced remarkably well when the GCM state is employed. These results indicate that the admixture of configurations with different shapes in the initial state leads to a substantially better agreement with the exact solution across both low- and high-energy regions.

\begin{figure}[]
\centering 
\includegraphics[width=8.5cm]{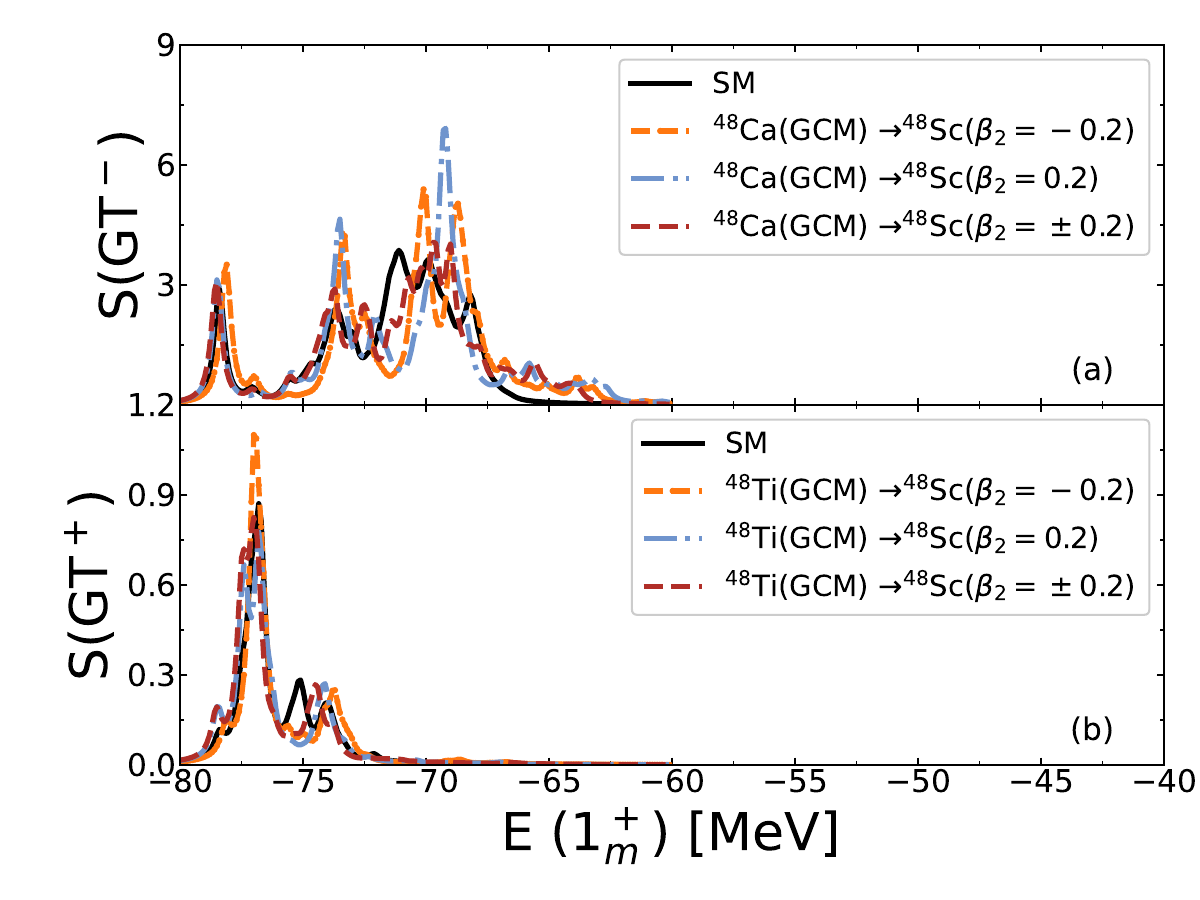}   
\caption{ (Color online)  The distribution of GT transition strength  for (a)  $^{48}$Ca and (b) $^{48}$Ti as a  function of the energy of the $1^+$ states in $^{48}$Sc from different calculations.  }
\label{fig:Ti48-sum-rule-compare} 
\end{figure} 

Figure~\ref{fig:Ti48-sum-rule-compare} shows the GT transition strengths for $^{48}$Ca and $^{48}$Ti obtained with GCM wave functions for the ground states of both nuclei. For the odd-odd nucleus $^{48}$Sc,  we restrict our analysis to cases using three different wave functions: the configuration of oblate energy minimum ($\beta_2=-0.2$), that of the prolate energy minimum ($\beta_2=+0.2$), and their admixture, respectively. In general, including the admixture of oblate and prolate deformed configurations improves the description of the GT transitions in both nuclei. Relative to calculations with only the prolate configuration, adding the oblate component yields a modest improvement for transitions to low-lying states but leads to a substantial enhancement for transitions to high-lying states. Nevertheless, the height of the first GT peak, around $E(1^+_m)=-78.4$ MeV, for the transition from $^{48}$Ti to $^{48}$Sc remains significantly overestimated.

 \begin{figure}[]
\centering 
\includegraphics[width=8.5cm]{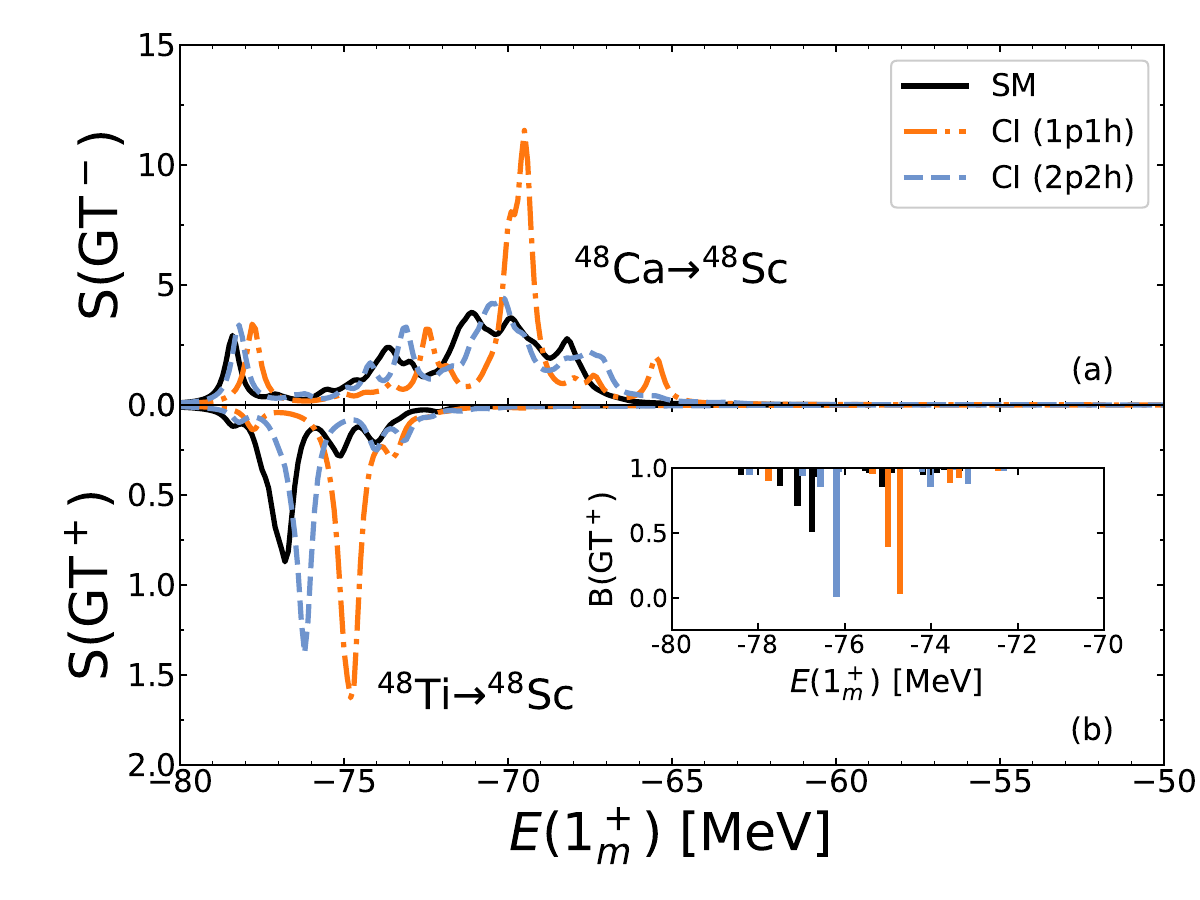}   
\caption{ (Color) The distributions of ${\rm GT}^-$ transition from  $^{48}$Ca (a) and ${\rm GT}^+$ transition from $^{48}$Ti (b) as a  function of the energy of the $1^+$ states in $^{48}$Sc, obtained from CI(1p1h), CI(2p2h) and shell-model calculations.}
\label{fig:48Ca+Ti} 
\end{figure}

Figure~\ref{fig:48Ca+Ti} compares the results of the CI(1p1h) and CI(2p2h) calculations, providing insight into the impact of model-space truncation on GT transitions. For the GT$^-$ transition in $^{48}$Ca, extending the CI model space from 1p1h to 2p2h shifts the first peak to lower excitation energy, substantially reduces the height of the main peak, broadens its width, and brings the results into closer agreement with the exact shell-model calculation. For the GT$^+$ transition in $^{48}$Ti, the main peak is likewise shifted to lower energy and its height is substantially reduced, although this improvement remains insufficient to fully reproduce the exact result. Comparing the PGCM results in Fig.~\ref{fig:Ti48-sum-rule-compare} with the CI(2p2h) results in Fig.~\ref{fig:48Ca+Ti}, one finds that the PGCM exhibits overall better agreement with the exact shell-model result.

\begin{figure}[] 
  \centering  
\includegraphics[width=8.5cm]{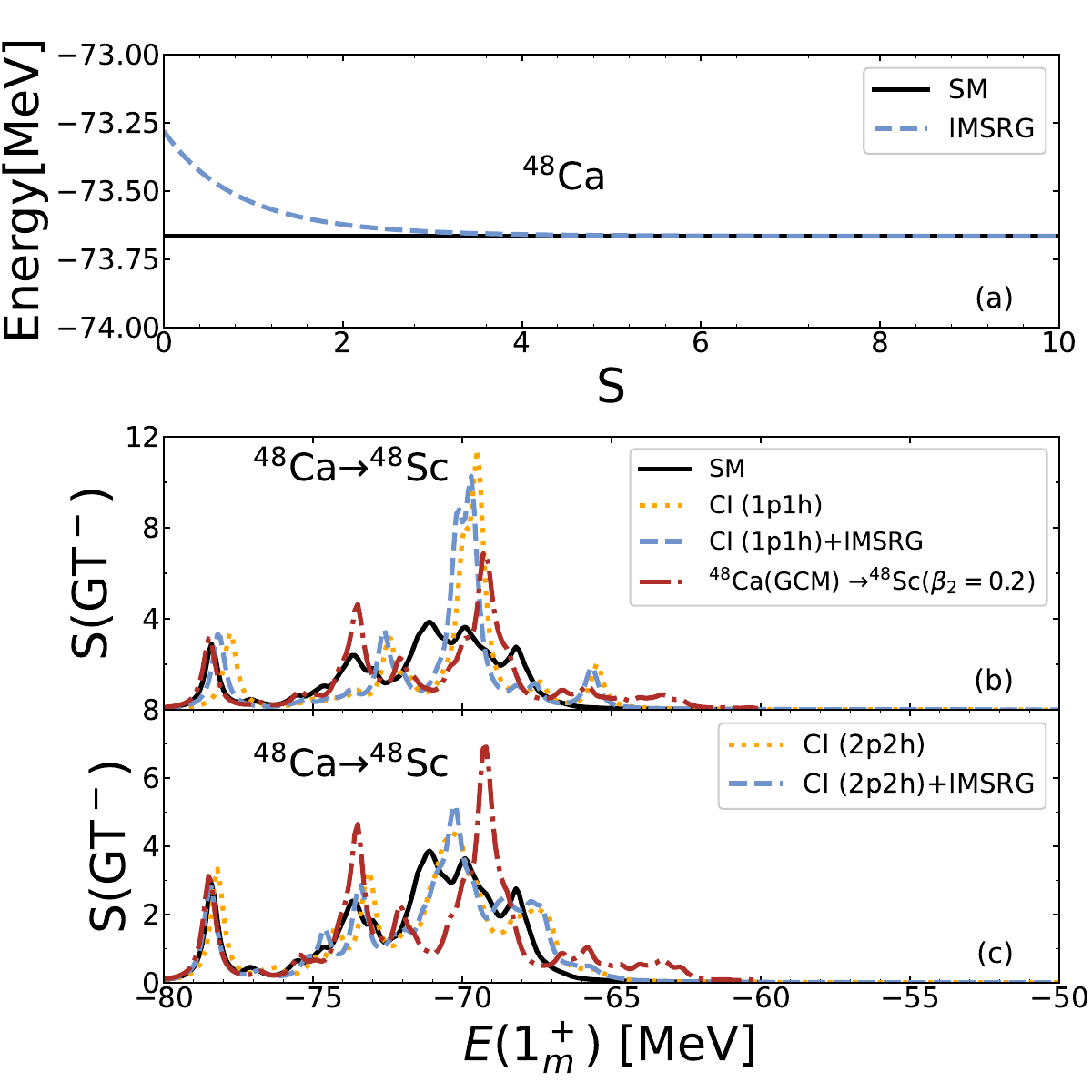}   
\caption{(Color online)  (a) The ground-state energy of $^{48}$Ca as a function of the flow parameter $s$ in the IMSRG calculation. (b) The distribution of ${\rm GT}^-$ transition strength $B({\rm GT}^-: 0^+_1 \to 1^+_m)$ from $^{48}$Ca to $^{48}$Sc as a  function of the energy of the $1^+$ states in $^{48}$Sc in the CI(1p1h) calculation with or without IMSRG. (c) The same as (b) but in the CI(2p2h) calculation.}
\label{fig:48Ca-IMSRG} 
\end{figure} 

The in-medium similarity renormalization group (IMSRG) method~\cite{Hergert:2016PR} has achieved remarkable success
in the {\em ab initio} studies of atomic nuclei starting from chiral Hamiltonians. Following Ref.~\cite{Yao:2018PRC}, we solve the IMSRG for $^{48}$Ca starting from the shell-model interaction GXPF1A based on the normal-ordered two-body (NO2B) approximation. Both Hamiltonian and GT operator are consistently evolved for all comparisons. The ground state of $^{48}$Ca  as a function of the flow parameter $s$ is displayed in Fig.~\ref{fig:48Ca-IMSRG}(a). It is shown that the energy converges toward the exact solution obtained by shell-model diagonalization. Using the evolved shell-model Hamiltonian, we perform CI(1p1h) and CI(2p2h) calculations for the GT transition strengths from $^{48}$Ca to $^{48}$Sc, employing the corresponding evolved GT transition operators. For comparison, results obtained with the unevolved operators are also presented. As seen in Fig.~\ref{fig:48Ca-IMSRG}(b), the IMSRG evolution leads to a modest improvement in the CI(1p1h) description; nevertheless, its performance remains slightly inferior to that of the PGCM. In contrast, the CI(2p2h) calculations with IMSRG exhibit slightly better agreement with the exact solutions than the PGCM. These observations suggest that combining the IMSRG with the PGCM could further enhance the description of GT transitions, an extension that we leave for future work.

\subsection{The NME of the 2$\nu\beta\beta$ decay in $^{48}$Ca}

The NME of the $2\nu\beta\beta$ decay from the ground state of $^{48}$Ca to that of $^{48}$Ti is given by~\cite{Yao:2022PPNP}
   \beq \label{Eq:M2nbb}
   M^{2\nu} =  \sum_m\dfrac{\langle 0^+_f\vert\vert \sigma\tau^- \vert\vert 1^+_m\rangle
\langle 1^+_m\vert\vert\sigma\tau^- \vert\vert 0^+_i\rangle}{E(1^+_m) - [E(0^+_i)+E(0^+_f)]/2}
   \eeq
where $E(1^+_m)$, $E(0^+_i)$, and $E(0^+_f)$ are the energies of the $m$-th $1^+$ state of the intermediate odd-odd nucleus $^{48}$Sc, and the ground-state energies of $^{48}$Ca and $^{48}$Ti, respectively.

\begin{figure}[]
\centering 
\includegraphics[width=8.5cm]{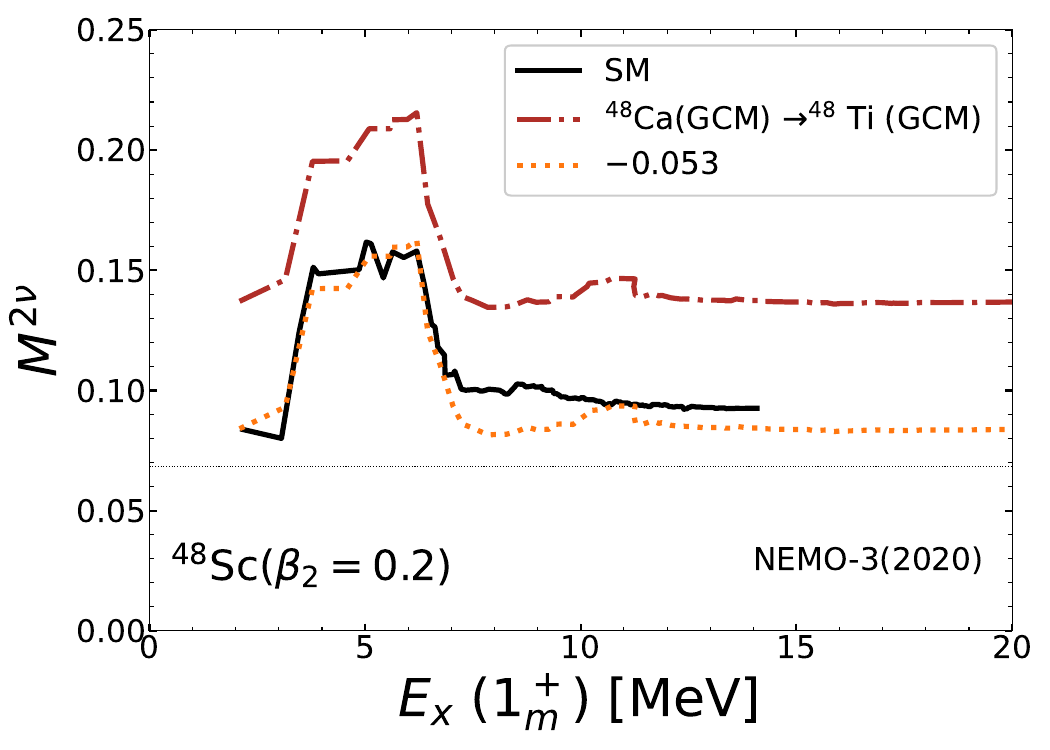}  
\caption{(Color online)  Cumulative NME $M^{2\nu}$ of $2\nu\beta\beta$ decay as a function of the excitation energy of the $1^+$ states in $^{48}$Sc from different calculations. The experiment data is taken from Ref.\cite{Barabash:2020nck}.}
\label{fig:2nbb-matrix-element} 
\end{figure} 

Using the relation
\begin{align}
Q_{\beta\beta}
&= E(0^+_i)-E(0^+_f)+2(m_n-m_p-m_e)\nonumber\\
&= E(0^+_i)-E(0^+_f)+1.56 ~{\rm (MeV)},
\end{align}
one obtains
\begin{align}
[E(0^+_i)+E(0^+_f)]/2
&= E(0^+_i)- [E(0^+_i)-E(0^+_f)]/2\nonumber\\
&= E(0^+_i)-(Q_{\beta\beta}-1.56)/2~{\rm (MeV)}.
\end{align}
The energy denominator entering the $2\nu\beta\beta$ matrix element can then be written as
\begin{align}
&E(1^+_m)-[E(0^+_i)+E(0^+_f)]/2\nonumber\\
&= E(1^+_m)-E(0^+_i)+ (Q_{\beta\beta}-1.56)/2 \nonumber\\
&= E_x(1^+_m)+E(6^+_1)-E(0^+_i)+(Q_{\beta\beta}-1.56)/2 ~{\rm (MeV)},
\end{align}
where $E_x(1^+_m)$ denotes the excitation energy of the $1^+_m$ state with respect to the ground state of $^{48}$Sc, which has spin-parity $6^+_1$. In practice, the entire $^{48}$Sc spectrum obtained from nuclear model calculations is shifted by aligning the lowest-lying $1^+_1$ state with its experimental data. Experimentally, the energy difference between the $^{48}$Sc ground state and the initial nucleus $^{48}$Ca is $E(6^+_1)-E(0^+_i)=0.503$~MeV. Substituting this value, the energy denominator becomes
\begin{align}
E(1^+_m) - [E(0^+_i)+E(0^+_f)]/2
&= E_x(1^+_m)+Q_{\beta\beta}/2-0.277 \nonumber\\
&= E_x(1^+_m)+1.857~{\rm (MeV)},
\end{align}
where the experimental value $Q_{\beta\beta}=4.2682$~MeV has been used.

Figure~\ref{fig:2nbb-matrix-element} shows the cumulated NME as a function of the excitation energy of the $1^+$ states in $^{48}$Sc, comparing the PGCM results with the exact shell-model calculation. The effective NME $M^{2\nu}_{\rm eff}$ can be derived from the half-life $T^{2\nu}_{1/2}=
5.3^{+1.2}_{-0.8}\times 10^{19}\ \mathrm{yr}$ based on the following formula~\cite{Barabash:2020nck}, 
\beq
[T^{2\nu}_{1/2}]^{-1} 
=G_{2\nu} |M^{2\nu}_{\rm eff}|^2
\eeq 
where $G_{2\nu}=1.555\times 10^{-17}\mathrm{yr}^{-1}$~\cite{Kotila:2012} is the phase-space factor. One finds $M^{2\nu}_{\rm eff}=0.035\equiv g^2_A m_ec^2 M^{2\nu}$~\cite{Barabash:2020nck}, where $m_e$ denotes the electron mass. Introducing a quenching factor  $q\simeq 0.78$ for the axial-vector coupling constant $g_A$ such that $qg_A\simeq 1.0$, the corresponding value of  $M^{2\nu}$ is obtained as 
\beq 
M^{2\nu} = M^{2\nu}_{\rm eff}/(q^2 g^2_A m_e)\simeq 0.0685~{\rm MeV}^{-1}.
\eeq 
This value is shown in Fig.\ref{fig:2nbb-matrix-element} for comparison. 

It is seen from Fig.\ref{fig:2nbb-matrix-element}  that most of the positive contribution to the  total NME $M^{2\nu}$ of $2\nu\beta\beta$ decay comes from two of the five intermediate $1^+$ states below 5 MeV excitation in $^{48}$Sc and the contribution of the high-lying intermediate states is not coherent, as  discussed in Ref. \cite{Horoi:2007}. This phenomenon is shown in the results of both exact shell-model and PGCM. The final NME $M^{2\nu}$ from the shell-model calculation is about 0.090 MeV$^{-1}$\cite{Horoi:2007} \footnote{Note that the value of $M^{2\nu}_{\rm eff}$ does not depend on the choice of the $g_A$ quenching factor $q$, while the $M^{2\nu}$ does. For example, the $M^{2\nu}$ by the shell model calculation based on Eq.(\ref{Eq:M2nbb}) is $0.0539/q^2\simeq0.090$ MeV$^{-1}$.}, about 30\% larger than the data of 0.0685 MeV$^{-1}$. Compared to the result by the shell-model calculation, the cumulated NME by the PGCM displays an almost constant offset of approximately $-0.053$ (corresponding to approximately 57\% of the total value), which originates primarily from the reduced matrix element $\langle {}^{48}\mathrm{Ti}(0^+_1)\Vert \sigma\tau^- \Vert {}^{48}\mathrm{Sc}(1^+_1)\rangle$, as discussed in Fig.~\ref{fig:Ti48-sum-rule-compare}. In the PGCM calculation, the reduced matrix elements are $\langle {}^{48}\mathrm{Ti}(0^+_1)\Vert \sigma\tau^- \Vert {}^{48}\mathrm{Sc}(1^+_1)\rangle = 0.345$ and $\langle {}^{48}\mathrm{Sc}(1^+_1)\Vert \sigma\tau^- \Vert {}^{48}\mathrm{Ca}(0^+_1)\rangle = 1.568$, compared with the corresponding shell-model values of $0.241$ and $1.496$, respectively. Consequently, the initial value of the $2\nu\beta\beta$ matrix-element curve is governed by the product $\langle 0^+_1\Vert \sigma\tau^- \Vert 1^+_1\rangle \langle 1^+_1\Vert \sigma\tau^- \Vert 0^+_1\rangle$, divided by the energy denominator, as given in Eq.~(\ref{Eq:M2nbb}).

\section{Summary}
 \label{sec:summary}
This work presents a minimal extension of the quantum-number projected generator coordinate method (PGCM) to describe Gamow–Teller (GT) transition strengths and the nuclear matrix elements (NMEs) for the two-neutrino double-beta ($2\nu\beta\beta$) decay  in even–even nuclei. By constructing states of odd-odd nuclei from quantum-number projected two quasiparticle excitations built on a HFB reference state, the approach enables a unified description of intermediate states in odd-odd nuclei within the PGCM framework. Benchmark calculations using a  valence-space ($fp$) shell-model Hamiltonian demonstrate that the method reasonably reproduces the GT transitions to the low-lying  and giant resonance states in calcium and titanium isotopes, with an overall performance comparable to, in some cases better than, configuration-interaction calculations truncated at the two-particle–two-hole level.  The NME of $2\nu\beta\beta$ decay from $^{48}$Ca to $^{48}$Ti is also evaluated without invoking the closure approximation. The result is overestimated by approximately 57\%, primarily due to the overestimated GT transition strength to the lowest-lying intermediate state. These findings establish the PGCM as a viable framework for describing $\beta$-decay observables in nuclei near closed shells and point to systematic improvements, such as extending the generator-coordinate space and incorporating the in-medium similarity renormalization group, that are expected to enhance its predictive power for more complex systems.    

\begin{acknowledgments} 
We thank R. Wirth and H. Hergert for helpful discussions in the early stage of this work. This work was supported in part by the National Natural Science Foundation of China (Grant Nos. 12405143, 12375119, and 12141501) and the Guangdong Basic and Applied Basic Research Foundation (Grant No. 2023A1515010936). 
\end{acknowledgments}


%

  \end{document}